\documentclass[floats,floatfix,amssymb,amsmath,aps,11pt,prd,superscriptaddress,nofootinbib,preprintnumbers,singlecolumn]{revtex4-1}
\usepackage{color}
\usepackage{subcaption}
\usepackage{graphicx}
\usepackage{hyperref}
\usepackage{appendix}
\usepackage{physics}
\usepackage{slashed}

\begin{document}

\title{A unified spin-harmonic framework for correlating pulsar timing, astrometric deflection, and shimmering gravitational wave observations}

\author{Giorgio Mentasti}
\email{g.mentasti21@imperial.ac.uk}
\thanks{ORCID: \href{https://orcid.org/0000-0003-1115-9220}{0000-0003-1115-9220}}
\affiliation{Department of Physics, Imperial College London, SW7 2AZ, London, United Kingdom}

\author{Carlo R. Contaldi}
\email{c.contaldi@imperial.ac.uk}
\thanks{ORCID: \href{https://orcid.org/0000-0001-7285-0707}{0000-0001-7285-0707}}
\affiliation{Department of Physics, Imperial College London, SW7 2AZ, London, United Kingdom}

\date{\today}
\begin{abstract}
\noindent We present a unified spin-weighted harmonic framework that delivers analytic, diagonal expressions for the overlap (correlation) functions of three low frequency gravitational wave observables-pulsar timing redshifts, astrometric deflections, and time-dependent image distortions (``shimmering''). Writing each response in spin-$s$ spherical harmonics and rotating to a basis in which the wave tensor has definite helicity, we obtain compact closed-form series for every auto- and cross-correlation, recovering the Hellings-Downs curve as the $s=0$ limit and deriving its astrometric ($s=\pm 1$) and shimmering ($s=\pm 2$) analogues. The formalism naturally extends to non-standard scalar-breathing, longitudinal, and vector polarisation modes, clarifying when higher-spin observables are (and are not) sourced and providing a complete set of harmonic spectra $C_\ell$ ready for parameter estimation pipelines. These results supply the common theoretical language needed to combine upcoming pulsar timing, Gaia-class astrometric, and high resolution imaging data sets, enabling coherent, multi probe searches for stochastic gravitational wave backgrounds, tests of general relativity and its alternatives across the nano- to micro-hertz gravitational wave band.
\end{abstract}

\maketitle


\section{Introduction}

The detection of gravitational waves (GWs) across an expanding range of frequencies continues to deepen our understanding of the universe and its most energetic phenomena. Ground-based interferometers have opened the high-frequency window around $f \sim 100$ Hz, revealing the mergers of stellar-mass compact binaries \cite{PhysRevLett.116.061102}. At lower frequencies, the upcoming space-based mission LISA will access the millihertz regime ($ f\sim 10^{-3}$ Hz), enabling the study of a wide variety of astrophysical sources such as extreme mass ratio inspirals and galactic binaries \cite{Baker:2019nia}. Meanwhile, at even lower frequencies, recent Pulsar Timing Array (PTA) results have provided compelling evidence for a nanohertz stochastic gravitational wave background \cite{EPTA:2023fyk, NANOGrav:2023gor,refId0}.

PTAs detect GWs by monitoring timing residuals in a galactic network of millisecond pulsars, where passing waves induce characteristic correlations in the arrival times of pulsar signals. In parallel, astrometric methods have emerged as a complementary probe of low-frequency GWs, where the apparent angular displacements of distant sources are tracked over time \cite{2011PhRvD..83b4024B, Darling2018}. Observations from Gaia \cite{Gaia} and future surveys such as LSST \cite{LSST:2008ijt} will enable unprecedented astrometric precision for a large number of observed astronomical objects, facilitating this form of gravitational wave detection on decade-long timescales.

Recently \cite{Mentasti:2024fgt}, a third method has been proposed, based on the shape distortions of resolved images of distant sources, as an inherently tensorial observable sensitive to gravitational waves. This effect, which the authors referred to as {\sl shimmering}, describes the time-dependent distortion of the apparent shape of astrophysical sources caused by gravitational waves. Unlike traditional cosmic shear in the framework of weak lensing, which is effectively time-independent on observational timescales, this shimmering effect is explicitly dynamical and is induced by propagating tensor modes. This new observable offers a conceptually distinct route to gravitational wave detection and provides a novel counterpart to PTA and astrometric techniques.

In this {\sl paper}, we extend and generalise the formalism first introduced in \cite{allen2024pulsar} and employed to the shimmering observable in \cite{Mentasti:2024fgt} to include not only the redshift and shimmering effects, but also astrometric observations within a unified framework. This formalism allows for the computation of the response functions of any scalar and tensorial observables to gravitational wave backgrounds with generalised polarisations. It yields, in a straightforward manner, characterisation of the overlap functions for any combination of observables, and the generalisations of the well-known Hellings–Downs (HD) curve \cite{HD}.

A key advantage of this approach lies in its ability to express the overlap functions for all three observables — PTA timing residuals, astrometric deflections, and shimmering distortions — through compact and structurally similar expressions. This reveals a deep commonality among these methods in their response to stochastic gravitational wave backgrounds. We further apply this formalism to gravitational waves beyond general relativity, characterising the response to alternative polarisation modes in a model-independent way \cite{PhysRevD.97.124058}.

By unifying these techniques under the same mathematical framework, this work not only clarifies the geometrical and physical connections among different low-frequency GW observables but also facilitates systematic comparisons of their sensitivities and polarisation dependencies.

The structure of the paper is as follows. In Section~\ref{sec:defs} we introduce the notation and recall the expressions for the antenna pattern of redshift, astrometric and \textit{shimmering} measurements. In Section~\ref{sec:Eresponse} we outline the general formalism and apply it step by step to compute the overlap functions for redshift, astrometric and \textit{shimmering} observables in the presence of a background of Einsteinian-polarised gravitational waves. In Section~\ref{sec:NonEresponse} we summarise the results of the same computation for the case of non-Einsteinian polarisations. We discuss our results in Section~\ref{sec:discussion}, where we also summarise the whole set of angular spectra we have obtained for quick reference.

\section{Redshift and astrometric effects of a GW}\label{sec:defs}

A photon emitted by an astrophysical source with four-momentum $p_\mu$ travels along null geodesics in a spacetime that is only approximately homogeneous and isotropic. It reaches the observer with a perturbed four-momentum $\tilde p_\mu$, which encodes both the background cosmological expansion and the intervening metric fluctuations. These effects manifest as an energy (redshift) change, $p_0 \!\to\! \tilde p_0$, and as an astrometric deflection of the propagation direction, $\mathbf p \!\to\! \tilde{\mathbf p}$.

Scalar perturbations sourced by non-relativistic matter generate redshift and lensing distortions that are effectively time-independent on accessible observational time-scales. In contrast, tensor perturbations—gravitational waves—produce genuinely time-dependent redshift and astrometric signatures. This time-dependence is the defining characteristic of this class of observables in which sources are monitored over long periods, of the order of years, to detect and characterise the spectral domain of the signal. Alongside this, the characterisation of the angular correlation of the time-dependent signal also serves a critical role in confirming any detection as tensorial in nature. The particular angular correlation that GWs induce in timing residuals was derived by \cite{HD} and is known as the HD curve. The claimed detection of an HD correlation for PTA observations has been an important milestone in verifying the claimed detections \cite{EPTA:2023fyk, NANOGrav:2023gor,refId0}. Astrometric and shimmering observables sourced by GWs also display angular correlations that are analogues to the HD curve and will be just as important in distinguishing any time-dependent signal from systematic effects.

The angular response in astrometric and redshift observables has been well studied for point sources (pulsars and stellar positions, see, e.g.~\cite{HD,2011PhRvD..83b4024B,OBeirne:2018slh,PhysRevD.97.124058,PhysRevD.105.063502}) and was recently extended to extended sources in~\cite{Mentasti:2024fgt}. Here we summarise the main results.
 
We recall that, in the long Earth-pulsar distance limit, a gravitational wave coming from the direction $\hat q$ induces an observed redshift modulation on a pulsar located in the direction $\hat n$ \cite{HD}
\begin{align}\label{eq:z_def}
z(\hat q,\hat n) = \frac{1}{2} \frac{h_{jk}(\hat q)\hat n^j\hat n^k}{1+\hat q_l\,\hat n^l}\,,
\end{align}
where direction unit vectors $\hat q$ and $\hat n$ are defined using spherical polar coordinates
\begin{align}
\hat{q}=\begin{pmatrix}
    \sin\theta\cos\phi\\
    \sin\theta\sin\phi\\
    \cos\theta
\end{pmatrix}\,,\qquad
\hat{n}=\begin{pmatrix}
    \sin\theta_s\cos\phi_s\\
    \sin\theta_s\sin\phi_s\\
    \cos\theta_s
\end{pmatrix}\,.
\end{align}
We use the subscript $s$ to denote the coordinates of the source.

Similarly, the astrometric deflection $\delta \hat n$ of a line of sight along direction $\hat n$ resulting from a gravitational wave $h_{ij}(\hat q)$ with propagation momentum aligned as $\hat k=-\hat q$  \cite{2011PhRvD..83b4024B,PhysRevD.97.124058}
\begin{align}\label{eq:delta_n}
\delta \hat n^i (\hat q,\hat n) = \frac{1}{2}\left[ \frac{\hat n^i+\hat q^i}{1+\hat q_l\,\hat n^l}h_{jk}(\hat q)\hat n^j\hat n^k - h^{i}_{\,j}(\hat q)\hat n^j\right]\,,
\end{align}
while the \textit{shimmering} effect on an extended object introduced in \cite{Mentasti:2024fgt} is encoded in the distortion matrix
\begin{align}\label{eq:distortion}
\widetilde \psi_{ij}(\hat q,\hat n)\equiv\frac{\partial \delta \hat n_i}{\partial \hat n^j} &= \frac{1}{2}\left[\delta_{ij}- \frac{\hat n_i+\hat q_i}{1+\hat q_l\,\hat n^l}\hat q_j\right] \frac{h_{rs}\hat n^r\hat n^s}{1+\hat q_l\, \hat n^l}+\frac{\hat n_i+\hat q_i}{1+\hat q_l\, \hat n^l}h_{jr}\hat n^r -\frac{1}{2} h_{ij}\,.
\end{align}
The angular dependence of a gravitational wave can be decomposed into all the possible polarisations
\begin{align}
h_{ij}(q)=\sum_\lambda h^\lambda e^\lambda_{ij}(q)\,,
\end{align}
with $\lambda$ running over any number of polarisations and the polarisation tensors $e^\lambda_{ij}(q)$ defined with respect to the unit vector aligned with the propagation and its tangent unit vectors. We follow the notation of \cite{PhysRevD.97.124058} in defining them. In standard GR, under diffeomorphism invariance and after imposing constraints and gauge choices, linearised gravity only allows two helicity states. These are known as ``Einsteinian'' or ``tensor'' polarisations
\begin{align}
e^+_{ij}(\hat q)&=\hat e^\theta_i\hat e^\theta_j-\hat e^\phi_i\hat e^\phi_j\,,\nonumber\\
e^{\cross}_{ij}(\hat q)&=\hat e^\phi_i\hat e^\theta_j+\hat e^\theta_i\hat e^\phi_j\,,\end{align}
with
\begin{align}
\hat q&=\{\sin\theta\cos\phi,\sin\theta\sin\phi,\cos\theta\}\,,\nonumber\\
e^\theta(\hat q)&=\{\cos\theta\cos\phi,\cos\theta\sin\phi,-\sin\theta\}\,,\nonumber\\
e^\phi(\hat q)&=\{-\sin\phi,\cos\phi,0\}\,.
\end{align}
Other theories of gravity where diffeomorphisms are broken or additional Lorentz‐covariant fields are introduced allow for scalar and vector propagating modes. The scalar polarisations can be represented as ``breathing'' and longitudinal modes
\begin{align}
e^S_{ij}(\hat q)&=\hat e^\phi_i\hat e^\phi_j+\hat e^\theta_i\hat e^\theta_j\,,\nonumber\\
e^L_{ij}(\hat q)&=\hat q_i\hat q_j\,,
\end{align}
and the two vector polarisations are
\begin{align}
e^x_{ij}(\hat q)&=\hat q_i\hat e^\theta_j+\hat e^\theta_i\hat q_j\,,\nonumber\\
e^y_{ij}(\hat q)&=\hat q_i\hat e^\phi_j+\hat e^\phi_i\hat q_j\,.
\end{align}
The scalar and vector polarisations are often referred to as ``non-Einsteinian'' modes.
When considering the delfections $\delta \hat n$ vector or distortion tensor $\tilde \psi$ in a direction $\hat n$ it is often useful to decompose quantities onto a tangential basis defined by the geodesic on the two sphere connecting the direction $\hat n$ and the GW propagation direction $\hat q$. as done in \cite{Mentasti:2024fgt}, we define the basis vectors as
\begin{align}\label{eq:ephietheta_obj}
\hat b_\phi(\hat q,\hat n)&=\frac{\hat n \times \hat q}{\sqrt{1- (\hat n\cdot\hat q)^2}}\,,\nonumber\\
\hat b_\theta(\hat q,\hat n)&=\frac{\hat n \times \hat b_\phi(\hat q,\hat n)}{\sqrt{1- (\hat n\cdot\hat b_\phi)^2}}\,.
\end{align}
This basis can then be used to define the two astrometric spin-$\pm$1 components
\begin{align}\label{eq:Fpm_def}
F_{\,_{\pm}\delta}(\hat q,\hat n)=\delta \hat n_i(\hat q,\hat n)\left(\hat b^i_\theta(\hat q,\hat n)\pm i\,\hat b^i_\phi(\hat q,\hat n)\right)\,,
\end{align}
and the four \textit{shimmering} ones as in \cite{Mentasti:2024fgt}
\begin{align}\label{eq:psiab_def}
\psi_{ab}(\hat q,\hat n) = \hat b^i_a(\hat q,\hat n)
\,\widetilde \psi_{ij}(\hat q,\hat n)\,\hat b^j_a(\hat q,\hat n)\equiv\begin{pmatrix}
-F_{\kappa}(\hat q,\hat n) -F_{\gamma_1}(\hat q,\hat n) & -F_{\gamma_2}(\hat q,\hat n)+F_{\omega}(\hat q,\hat n) \\
-F_{\gamma_2}(\hat q,\hat n)-F_{\omega}(\hat q,\hat n) & -F_{\kappa}(\hat q,\hat n) +F_{\gamma_1}(\hat q,\hat n)
\end{pmatrix}\,.
\end{align}
For convenience, we use the coefficients $\gamma_1$ and $\gamma_2$, to introduce the explicit spin-$\pm$2 quantities
\begin{align}
F_{\,_{\pm}\gamma}(\hat q,\hat n)=F_{\gamma_1}(\hat q,\hat n)\pm i\,F_{\gamma_2}(\hat q,\hat n)\,.
\end{align}

\section{Einsteinian response, spectral functions, and overlap functions}\label{sec:Eresponse}

In this section, we derive the diagonal form of the redshift, astrometric, and shimmering response functions for a left-handed gravitational wave. We summarise the steps of the derivation, which follow the diagonal formulation introduced in \cite{allen2024pulsar} in the case of the redshift response function. We then promote the formalism to compute the response functions for the newly defined astrometric observables, and finally, we summarise the steps of \cite{Mentasti:2024fgt} for computing the \textit{shimmering} observables.
The response functions to a left-handed gravitational wave will be employed in the following section to compute the overlap functions between different observables.

As in \cite{allen2024pulsar} we define $\hat v(\hat q)=\hat e^\theta(\hat q)+i\hat e^\phi(\hat q)$, and we obtain the explicit form of the observables introduced in \eqref{eq:z_def}, \eqref{eq:Fpm_def}, and \eqref{eq:psiab_def} for a purely left-handed polarized gravitational wave $h^E_{ij}(\hat q)=h^{+}_{ij}(\hat q)+i\,h^{\times}_{ij}(\hat q)$:
\begin{align}\label{eq:distortion_E}
z^E(\hat q,\hat n) &= \frac{1}{2} \frac{\left[\hat n\cdot \hat v(\hat q)\right]^2}{1+\hat q_l\,\hat n^l} \,,\nonumber\\
\delta \hat n^E_i (\hat q,\hat n) &= \frac{1}{2}\left[ \frac{\hat n_i+\hat q_i}{1+\hat q_l\,\hat n^l}\left[\hat n\cdot \hat v(\hat q)\right]^2 - \left[\hat n\cdot \hat v(\hat q)\right]\hat v_i(\hat q)\right]\,,\\
\widetilde \psi_{ij}^E (\hat q,\hat n)&= \frac{1}{2}\left[\delta_{ij}- \frac{\hat n_i+\hat q_i}{1+\hat q_l\,n^l}\hat q_j\right] \frac{\left[\hat n\cdot \hat v(\hat q)\right]^2}{1+\hat q_l\,\hat n^l}+\frac{\hat n_i+\hat q_i}{1+\hat q_l\,\hat n^l}\left[\hat n\cdot \hat v(\hat q)\right] \hat v_j(\hat q) -\frac{1}{2} \hat v_i(\hat q)\hat v_j(\hat q)\,,\nonumber
\end{align}
where the `$E$' superscript is used to denote the response functions to the transverse and traceless polarisations.
For convenience, we can evaluate the response functions for each of the observables in the reference system where the GW travels along the $z$-axis. We do this for all spin-$0$, spin-$\pm1$, and spin-$\pm2$ observables defined above
\begin{align}\label{eq:vars_Einst}
F^E_{z}(\hat z,\hat n)&=\frac{1}{2}e^{-2i(\phi-\phi_s)}(1-\cos\theta_s)\,,\nonumber\\
F^E_{\,_{+}\delta}(\hat z,\hat n)&=-e^{-2i(\phi-\phi_s)}\sin\theta_s\,,\nonumber\\
F^E_{\,_{+}\delta}(\hat z,\hat n)&=0\,,\nonumber\\
F^E_{\kappa}(\hat z,\hat n)&=-\frac{1}{2}e^{-2i(\phi-\phi_s)}(1-\cos\theta_s)\,,\\
F^E_{\omega}(\hat z,\hat n)&=i\,F_{\kappa}(\hat z,\hat n) \,,\nonumber\\
F^E_{{}\,_{-}\gamma}(\hat z,\hat n)&=e^{-2i(\phi-\phi_s)}\,,\nonumber\\
F^E_{{}_{+}\gamma}(\hat z,\hat n)&=0\,.\nonumber
\end{align}
The computation of the response function of the redshift and the \textit{shimmering} coefficients $\kappa,\omega$, which are scalar observables in the object's coordinates $\hat n$, proceeds as in the diagonal formalism introduced in \cite{allen2024pulsar}. For the astrometric and the shimmering observables, our definition of spin-weighted quantities allows us to extend the diagonal formalism to the higher helicity case as shown in \cite{Mentasti:2024fgt}. The fact that, under a suitable treatment of the rotation properties of the observables, the diagonal form of the correlation functions is also present in the spin-1 and spin-2 cases greatly simplifies the calculation of the overlap functions in the astrometric and shimmering cases.

To achieve this, one needs to perform an additional step, creating variables that transform as scalars by applying spin-raising and spin-lowering operators\footnote{We refer to \cite{Zaldarriaga:1996xe} for more details on the spin-raising/lowering operators.}
\begin{align}
\partial_s f(\theta, \phi) & =-\sin ^s(\theta)\left[\frac{\partial}{\partial \theta}+i \csc (\theta) \frac{\partial}{\partial \phi}\right] \sin ^{-s}(\theta){ }_s f(\theta, \phi)\,,\nonumber \\
\bar{\partial}_s f(\theta, \phi) & =-\sin ^{-s}(\theta)\left[\frac{\partial}{\partial \theta}-i \csc (\theta) \frac{\partial}{\partial \phi}\right] \sin ^s(\theta){ }_s f(\theta, \phi)\,,
\end{align}
where ${ }_s f(\theta, \phi)$ is a generic function of spin $s$.
In particular, the astrometric observables $F_{{}_{\pm} \delta}$ have spin $\mp 1$, while $F_{{}_{\pm} \gamma}$ have spin $\pm 1$. This introduces the spin-0 quantities
\begin{align}
F^E_{_{+}\xi}(\hat z,\hat n)&=\slashed\partial F^E_{{}_{+}\delta}(\hat z,\hat n)=2e^{-2i(\phi-\phi_s)}(\cos\theta_s-1)\,,\nonumber\\
F^E_{_{-}\zeta}(\hat z,\hat n)&=\slashed\partial^2F^E_{{}\,_{-}\gamma}(\hat z,\hat n)=2e^{-2i(\phi-\phi_s)}\frac{1-\cos\theta_s}{1+\cos\theta_s}\,.
\end{align}
The spin-$0$ quantities above are invariant under rotations of the frame and can therefore be used to define the invariant, diagonal spherical harmonic coefficients of the response functions
\begin{align}
F^E_{z}(\hat z,\hat n)&=\sum_{\ell=2}^\infty q^{z,E}_\ell Y^\star_{\ell -2}(\hat n)\,,\nonumber\\
F^E_{_{+}\xi}(\hat z,\hat n)&=\sum_{\ell=2}^\infty q^{\xi,E}_\ell Y^\star_{\ell -2}(\hat n)\,,\\
F^E_\kappa(\hat z,\hat n)&=\sum_{\ell=2}^\infty q^{\kappa,E}_\ell Y^\star_{\ell -2}(\hat n)\,,\nonumber\\
F^E_{_{-}\zeta}(\hat z,\hat n)&=\sum_{\ell=2}^\infty q^{\zeta,E}_\ell Y^\star_{\ell -2}(\hat n)\,,\nonumber
\end{align}
with
\begin{align}
q^{z,E}_\ell&=\left(-1\right)^{\ell}\sqrt{\frac{4\pi(2\ell+1)}{(\ell-1)\ell(\ell+1)(\ell+2)}}\,e^{-2i\phi}\,,\nonumber\\
q^{\xi,E}&=-4\,q^{z,E}_\ell\,,\\
q^{\kappa,E}&=q^{z,E}_\ell\,,\nonumber\\
q_\ell^{\zeta,S}&=2(\ell-1)(\ell+2)\,q^{z,E}_\ell\,.\nonumber
\end{align}
It can then be shown \cite{allen2024pulsar,Mentasti:2024fgt} that under rotation to a general reference system, where the GW is not aligned with the $z$-axis, and applying an inverse spin-raising or lowering step, we obtain the following response functions
\begin{align}\label{eq:single_source_diagform}
F^E_{z}(\hat q,\hat n)&=\sum_{\ell=2}^\infty A_\ell^{z,E}\,\sum_{m=-\ell}^\ell {}^{\,}_{+2}Y^{\,}_{\ell m}(\hat q)\,Y_{\ell m}^\star(\hat n)\,,\nonumber\\
F^E_{\,_{+}\delta}(\hat q,\hat n)&=\sum_{\ell=2}^\infty A_\ell^{\,_{+}\delta,E}\,\sum_{m=-\ell}^\ell {}^{\,}_{+2}Y^{\,}_{\ell m}(\hat q)\,{}^{\,{}}_{-1}Y_{\ell m}^\star(\hat n)\,,\nonumber\\
F^E_{\,_{+}\delta}(\hat q,\hat n)&=0\,,\nonumber\\
F^E_{\kappa}(\hat q,\hat n)&=\sum_{\ell=2}^\infty A_\ell^{\kappa,E}\,\sum_{m=-\ell}^\ell  {}^{\,}_{+2}Y^{\,}_{\ell m}(\hat q)\,Y_{\ell m}^\star(\hat n)\,,\\
F^E_{\omega}(\hat q,\hat n)&=\sum_{\ell=2}^\infty A_\ell^{\omega,E} \,\sum_{m=-\ell}^\ell {}^{\,}_{+2}Y^{\,}_{\ell m}(\hat q)\,Y_{\ell m}^\star(\hat n)\,,\nonumber\\
F^E_{{}^{\,}\,_{-}\gamma}(\hat q,\hat n)&=\sum_{\ell=2}^\infty A_\ell^{\,\,_{-}\gamma,E} \,\sum_{m=-\ell}^\ell {}^{\,}_{+2}Y^{\,}_{\ell m}(\hat q)\,{}^{\,{}}_{-2}Y_{\ell m}^\star(\hat n)\,,\nonumber\\
F^E_{{}^{\,}_{+}\gamma}(\hat q,\hat n)&=0\,,\nonumber
\end{align}
with
\begin{align}
A^{\,_{+}\delta,E}_\ell&=-(-1)^\ell\frac{16\pi}{\sqrt{(\ell+2)(\ell-1)}}\frac{1}{(\ell+1)\ell}\,,\nonumber\\
A_\ell^{z,E}&=A_\ell^{\kappa,E}=(-1)^\ell\, 4\pi \sqrt{\frac{(\ell-2)!}{(\ell+2)!}}\,,\\
A_\ell^{\omega,E}&=i(-1)^\ell\, 4\pi \sqrt{\frac{(\ell-2)!}{(\ell+2)!}}=i\,A_\ell^{\kappa,E}\,,\nonumber\\
A_\ell^{\,\,_{-}\gamma,E}&=(-1)^\ell \,8\pi\frac{(\ell-1)!}{(\ell+1)!}\,.\nonumber
\end{align}
For convenience, we introduce the factor
\begin{equation}
_s^{\,}{\cal N}_\ell = \sqrt{\frac{(\ell-s)!}{(\ell+s)!}}\,,
\end{equation}
to give the diagonal spectral functions
\begin{align}\label{eq:Als}
A^{\,_{+}\delta,E}_\ell&=-(-1)^\ell 16\pi \frac{_2^{\,}{\cal N}_\ell}{\sqrt{\ell(\ell+1)}}\,,\nonumber\\
A_\ell^{z,E}&=A_\ell^{\kappa,E}=(-1)^\ell\, 4\pi \; _2^{\,}{\cal N}_\ell\,,\\
A_\ell^{\omega,E}&\equiv i\,A_\ell^{\kappa,E}\,,\nonumber\\
A_\ell^{\,\,_{-}\gamma,E}&=(-1)^\ell \,8\pi\; _1^{\,}{\cal N}_\ell^2\,.\nonumber
\end{align}
Note that, in~\eqref{eq:single_source_diagform}, the sum over angular multiples $\ell$ starts at $\ell=2$ in all cases, including for observables with spin-weight less than 2. This is due to the presence of the spin-2 $_{2}Y_{\ell m}(\hat q)$ basis function that accounts for the rotation of the GW direction $\hat q$. This constrains the response function to have only quadrupolar or higher contributions, as expected for tensor polarisations. The second basis function will instead reflect the spin-weight of the observables around the direction $\hat n$ as it transforms under the same rotation. 

\subsection{Einsteinian Overlap functions}

The diagonal form obtained in \eqref{eq:single_source_diagform} allows us to derive the overlap functions between different observables in a few simple steps. We start by computing, in the Einsteinian case, the integrals of two observables; $A$ having spin-weight $s$ and $B$ having spin-weight $s'$.  Their generic diagonal form will be 
\begin{align} 
F_{A}^E(\hat q,\hat n)&=\sum_{\ell=2}^\infty A_\ell^{A,E}\sum_m \,_{2}Y_{\ell m}(\hat q)\,_{s}Y^\star_{\ell m}(\hat n)\,,\\
F_{B}^E(\hat q,\hat n)&=\sum_{\ell=2}^\infty A_\ell^{B,E}\sum_m \,_{2}Y_{\ell m}(\hat q)\,_{s'}Y^\star_{\ell m}(\hat n)\,.\nonumber
\end{align}
The correlation between observables can be computed by taking the angular integral
\begin{align}
\Gamma_{v,E}^{AB}(\hat n_1,\hat n_2) &\equiv \int d^2\hat q\, F^{E}_A(\hat q,\hat n_1)\left[F^{E}_B(\hat q,\hat n_2)\right]^\star\,,
\end{align}
which can be reduced further by introducing the Wigner rotation operator ${\cal D}^\ell_{ss'}$ and the related small-Wigner operator $d^\ell_{ss'}$, and introducing the cross-correlation angular power spectra $C_\ell^{AB} \equiv A_\ell^A A_\ell^{B\star}$, evaluates to
\begin{align}\label{eq:ECls}
\Gamma_{v,E}^{AB}(\hat n_1,\hat n_2) &=\left[\sum_{\ell m}C_\ell^{AB}\,_{s}Y_{\ell m}^\star(\hat n_1)\,_{s'}Y_{\ell m}(\hat n_2)\right]\,,\nonumber\\
&=(-1)^{-s}\sum_{\ell}\sqrt{\frac{2\ell+1}{4\pi}}\,C_\ell^{AB} \,_{-s}Y_{\ell s'}(\beta,\alpha)e^{is\epsilon}\,,\nonumber\\
&=\sum_{\ell}\frac{2\ell+1}{4\pi}\,C_\ell^{AB}\,\mathcal{D}_{s's}^{\ell\star}(\alpha,\beta,\epsilon)\,,\\
&=\frac{1}{4\pi}\sum_{\ell}\,(2\ell+1)\,C_\ell^{AB}\,d_{s's}^\ell(\beta) e^{-i(s'\alpha+s\epsilon)}\,,\nonumber 
\end{align}
where we used the addition theorem for spherical harmonics\footnote{See also \cite{allen2024pulsar} and \cite{PhysRevD.90.082001} for the technical details of the computation and the conventions used to introduce the Wigner-D symbols.} and where $(\alpha,\beta,\epsilon)$ are the three Euler angles defined by $\hat n_1,\hat n_2,\hat z$ as described in appendix A of \cite{Gair:2015hra}. In particular, in the reference system where the first object is located along the $z$-axis, one finds $\cos\beta=\hat n\cdot \hat n'$ and $\alpha=\epsilon=0$. The subscript $v$ indicates that these are the correlation functions for a {\sl left-handed} polarised GW, and we use this to distinguish them from those for unpolarised correlation functions. The two are trivially related in some cases, but in other cases, two of the polarised correlation functions must be combined to obtain the unpolarised version, as discussed below. To compute these correlators for our observables, we simply need to set $s=0$ for $z,\kappa,\omega$, then $s=-1$ for the $_{+}\delta$ and $s=-2$ for the $\,_{-}\gamma$ coefficients.

The presence of the phases in~\eqref{eq:ECls} follows from the fact that correlation functions between variables that do not transform trivially under rotations are not invariant under the same rotations. This is simply an artefact of the fact that the coordinate system used to define the higher helicity variables is not invariant; the angular correlation functions are frame dependent. However, the angular power spectra $C_\ell^{AB}$ are frame independent. For simplicity, the frame dependence can be ignored by considering solely the harmonic spectra or by only considering angular correlations for which the variables at either direction are rotated to the tangential basis aligned with the geodesic between the two directions being correlated~\cite{Ng:1997ez,10.1111/j.1365-2966.2004.07737.x,Contaldi:2015boa}. In this case, the phase sensitive term $\exp[-i(s'\alpha+s\epsilon)]\to 0$.

\begin{figure}[t]
  \centering
\includegraphics[width=10cm]{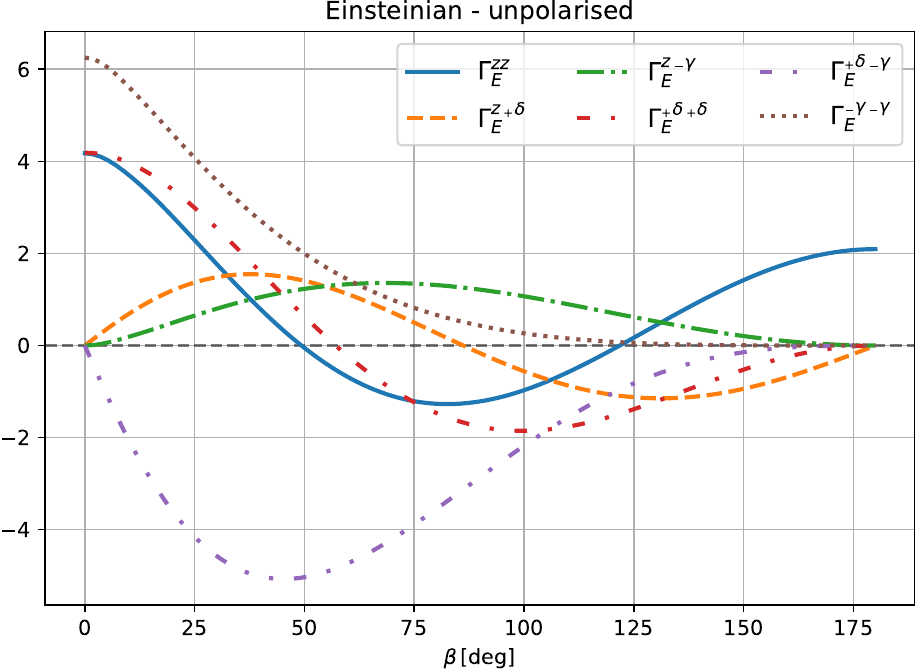}
  \caption{%
    Angular correlation functions for an Einsteinian‑unpolarised GW background, defined in Eq.~\eqref{eq:Gamma_unpol_def}, in the aligned frame as a function of the separation angle $\beta=\arccos(\hat n_1\!\cdot\!\hat n_2)$. We show the unique functions only. A lookup table of equivalent functions is shown in Table~\ref{tab:Einsteinian-symmetries}}
    \label{fig:Einsteinian}
\end{figure}

\begin{table}[]
    \centering
\begin{tabular}{|l | l |} 
     \hline
     Prototype & Equivalent overlap functions \\ [0.5ex] 
     \hline\hline
     $\Gamma^{zz}$ & $\Gamma^{\kappa\kappa}$, $\Gamma^{z\kappa}$, $\Gamma^{\kappa z}$, $\Gamma^{\omega\omega}$ \\ 
     \hline
     $\Gamma^{z_{+}\delta}$ & $\Gamma^{\kappa_{+}\delta}$, $-i\,\Gamma^{\omega_{+}\delta}$, $\Gamma^{{}_{-}\delta z}$, $\Gamma^{{}_{-}\delta\kappa}$, $-i\,\Gamma^{{}_{-}\delta\omega}$, $-\,\Gamma^{_{+}\delta z}$,  $-\,\Gamma^{\delta\kappa}$, $i\,\Gamma^{_{+}\delta \omega}$, $-\,\Gamma^{z_{-}\delta}$, $-\,\Gamma^{\kappa_{-}\delta}$, $i\,\Gamma^{\omega_{-}\delta}$\\
     \hline
     $\Gamma^{z \,_{-}\gamma}$ & $\Gamma^{\kappa_{-}\gamma}$, $-i\,\Gamma^{\omega_{-}\gamma}$, $\Gamma^{\gamma z}$, $\Gamma^{\gamma\kappa}$, $-i\,\Gamma^{\gamma\omega}$, $-\,\Gamma^{_{-}\gamma z}$,  $-\,\Gamma^{_{-}\gamma\kappa}$, $-i\,\Gamma^{_{-}\gamma \omega}$, $\,\Gamma^{z\gamma}$, $\,\Gamma^{\kappa\gamma}$, $-i\,\Gamma^{\omega\gamma}$ \\ 
     \hline
     $\Gamma^{\delta_{-}\gamma}$ & $\Gamma^{\gamma_{-}\delta}$, $-\Gamma^{_{-}\gamma_{+}\delta}$, $-\Gamma^{_{-}\delta_{-}\gamma}$ \\
     \hline
     $\Gamma^{_{+}\delta_{+}\delta}$ & $\Gamma^{_{-}\delta_{-}\delta}$ \\
     \hline
     $\Gamma^{_{-}\gamma\,_{-}\gamma}$ & $\Gamma^{\gamma\gamma}$ \\ [1ex] 
     \hline
    \end{tabular}
      \caption{Lookup table of equivalent correlation functions for all non‑zero combinations of Einsteinian overlap functions. The left column shows the unique functions shown in Fig.~\ref{fig:Einsteinian} and the right column shows all equivalent functions to each one. Note that, since some of our variables are complex, spin-weighted combinations of real observables, several of the correlation functions are imaginary.
  The remaining overlap functions, such as $\Gamma^{z\omega}_{E},\Gamma^{k\omega}_{E},\Gamma^{{}_{\pm}\delta _{\mp}\delta}_{E},\Gamma^{{}_{\pm}\gamma _{\mp}\gamma}_{E},\Gamma^{_{\pm}\delta _{\pm}\gamma}_{E}$ all vanish.}
    \label{tab:Einsteinian-symmetries}
\end{table}

Finally, we can the compute the general overlap functions $\Gamma^{XY}_{E}(\hat n_1,\hat n_2)$ for an unpolarized background (i.e., when the $+$ and the $\times$ polarizations have the same amplitude in the background)\footnote{The computation is the same for the case of a vectorially unpolarized GW background, i.e. when the GW amplitude of the $h^x$ and $h^y$ modes follow the same statistics. In that case, one can simply follow the same steps of the Einsteinian case, by just making the substitution $+\to x$ and $\times\to y$.}.
Some care is required at this point, as some of the observables considered (in our case, $_{\pm}\delta$ and $_{\pm}\gamma$) are complex quantities. Since we computed the overlap functions $\Gamma^{XY}_{v,E}(\hat n_1,\hat n_2)$ for a pure \textit{left} handed GW, which is also defined as a complex tensor, we need to proceed more carefully when we want to we compute correlators between two complex quantities.
In total generality, consider now the case where the two observables $A$ and $B$ are complex \footnote{for example, these could be a measurement of $_{+}\delta$ coefficient at object 1 and $_{+}\gamma$ at object 2.} and we wish to compute the unpolarized overlap functions. This can be achieved by forming the combination
\begin{align}\label{eq:Gamma_unpol_def}
\Gamma^{AB}_E(\hat n_1,\hat n_2)=\langle AB^\star\rangle_{\rm unpol}=\int d^2\hat q\left[A^{\,}_+(\hat n_1,\hat q)B^\star_+(\hat n_2,\hat q)+A^{\,}_\times(\hat n_1,\hat q)B^\star_\times(\hat n_2,\hat q)\right]\,,
\end{align}
where $A_{+/\times}(\hat n_1,\hat q)$ and $B_{+/\times}(\hat n_2,\hat q)$ are the response functions of the observable A and B to a purely $+$ and $\times$ polarised GW, respectively. As the GW tensors for the two polarisations $h_{ij}^+$ and $h_{ij}^\times$ are defined as real numbers, we find that the conjugate observables $\bar A$ and $\bar B$ for a pure $+$ and $\times$ polarised GW satisfy
\begin{align}
\bar A_{+/\times}(\hat n_1,\hat q)&=A^\star_{+/\times}(\hat n_1,\hat q)\,,\nonumber\\
\bar B_{+/\times}(\hat n_1,\hat q)&=B^\star_{+/\times}(\hat n_1,\hat q)\,.
\end{align}
In~\eqref{eq:ECls} we  computed the response functions for a purely left-handed polarised GW, such as
\begin{align}
\Gamma_{v,E}^{A,B}(\hat n_1,\hat n_2)=\langle A^{\,}_v B^\star_v\rangle&=\int d^2\hat q\,A_v(\hat n_1,q) B^\star_v(\hat n_2,\hat q)\,,\nonumber\\
A_v(\hat n_1,q)&=A_+(\hat n_1,q)+iA_\times(\hat n_1,q)\,,\nonumber\\
B_v(\hat n_1,q)&=B_+(\hat n_1,q)+iB_\times(\hat n_1,q)\,,
\end{align}
and similarly we can compute $\Gamma_{v,E}^{\bar A,B},\Gamma_{v,E}^{\bar A,\bar B}$ and $\Gamma_{v,E}^{A,\bar B}$. A direct substitution of these expressions in the following, making use of the conjugation property, shows that
\begin{align}\label{eq:gamma_unpol}
\Gamma_{E}^{A,B}(\hat n_1,\hat n_2)=\frac{1}{2}\left[\Gamma_{v,E}^{A,B}(\hat n_1,\hat n_2)+\Gamma_{v,E}^{\bar B,\bar A}(\hat n_2,\hat n_1)\right]\,.
\end{align}
For example, we have
\begin{align}
\Gamma^{_{+}\delta,\,_{-}\gamma}_E(\hat n_1,\hat n_2)=\frac{1}{2}\left[\Gamma^{_{+}\delta,\,_{-}\gamma}_{E,v}(\hat n_1,\hat n_2)+\Gamma^{_{+}\gamma,_{-}\delta}_{E,v}(\hat n_2,\hat n_1)\right]\,,
\end{align}
and so forth. The general formulae above are still valid when combining a real quantity with a complex one, or two real quantities, given that for a real quantity, we have $\bar A_{+/\times}=A_{+/\times}$. For example, we can compute
\begin{align}
\Gamma^{_{+}\delta,z}_E(\hat n_1,\hat n_2)&=\frac{1}{2}\left[\Gamma^{_{+}\delta,z}_{E,v}(\hat n_1,\hat n_2)+\Gamma^{z,_{-}\delta}_{E,v}(\hat n_2,\hat n_1)\right]\,,\nonumber\\
\Gamma^{\kappa,z}_E(\hat n_1,\hat n_2)&=\frac{1}{2}\left[\Gamma^{\kappa,z}_{E,v}(\hat n_1,\hat n_2)+\Gamma^{z,\kappa}_{E,v}(\hat n_2,\hat n_1)\right]\,.
\end{align}
In Fig.~\ref{fig:Einsteinian} we show the resulting correlation functions as a function of $\beta=\arccos(\hat n_1\cdot \hat n_2)$. As described above, we visualise the functions representing the choice of tangent basis aligned along the geodesic between the directions in which the relative phase factor vanishes. In this case, the Eq.~\eqref{eq:gamma_unpol} displayed in Fig.~\ref{fig:Einsteinian} takes the form
\begin{align}
\Gamma_{E}^{A,B}(\beta)&=\frac{1}{2}\left[\Gamma_{v,E}^{A,B}(\beta)+\Gamma_{v,E}^{\bar B,\bar A}(\beta)\right]\,,
\end{align}
where
\begin{align}
\Gamma_{v,E}^{AB}(\beta) &=\frac{1}{4\pi}\sum_{\ell}\,(2\ell+1)\,C_\ell^{AB}\,d_{s's}^\ell(\beta)\,.
\end{align}

\section{Non-Einsteinian polarisations}\label{sec:NonEresponse}

We now consider observable signatures generated by non-Einsteinian polarisations. Following a similar derivation, we can evaluate the response for each of the observables considered above.

\subsection{Scalar breathing mode response and spectral functions}

We start with the scalar breathing (S) mode defined by
\begin{equation}
h_{ij}^S(\hat q)=\frac{1}{2}\left[\hat v_i(\hat q)\hat v_j^\star(\hat q)+\hat v_i^\star(\hat q)\hat v_j(\hat q)\right]\,.
\end{equation}
The response functions analogous to those in~\eqref{eq:distortion_E} are
\begin{align}\label{eq:vars_S}
F^S_{z}(\hat z,\hat n)&=\frac{1}{2}(1-\cos\theta_s)\,,\nonumber\\
F^S_{\,_{\pm}\delta}(\hat z,\hat n)&=-\frac{1}{2}\sin\theta_s\,,\\
F^S_{\kappa}(\hat z,\hat n)&=\frac{1}{2}\cos\theta_s\,,\nonumber\\
F^S_{\omega}(\hat z,\hat n)&=0 \,,\nonumber\\
F^S_{{}_{\pm}\gamma}(\hat z,\hat n)&=\pm\frac{1}{2}\left(1-\cos\theta_s\right)\,.\nonumber
\end{align}
As before, we use raising and lowering operators to define spin-0 quantities from all higher spin astrometric and shimmering observables
\begin{align}
F^S_{_{-}\xi}(\hat z,\hat n)&=\slashed\partial F^S_{{}_{+}\delta}(\hat z,\hat n)=\cos\theta_s\,,\nonumber\\
F^S_{_{+}\xi}(\hat z,\hat n)&=\bar{\slashed\partial}F^S_{{}_{+}\delta}(\hat z,\hat n)=\cos\theta_s\,,\nonumber\\
F^S_{_{-}\zeta}(\hat z,\hat n)&=\slashed\partial^2F^S_{{}\,_{-}\gamma}(\hat z,\hat n)=\partial^2_\mu\left[(1-\mu^2)\left(-\frac{1}{2}\left(1-\mu\right)\right)\right]=\frac{1-3\cos\theta_s}{2}\,,\nonumber\\
F^S_{_{+}\zeta}(\hat z,\hat n)&=\bar{\slashed\partial}^2F^S_{{}_{+}\gamma}(\hat z,\hat n)=\partial^2_\mu\left[(1-\mu^2)\frac{1}{2}\left(1-\mu\right)\right]=\frac{3\cos\theta_s-1}{2}\,,
\end{align}
and expand them in spherical harmonics as in the Einsteinian case
\begin{align}
F^S_{z}(\hat z,\hat n)&=q^{z,S}_0 Y^\star_{0 0}(\hat n)+q^{z,S}_1 Y^\star_{1 0}(\hat n)\,,\nonumber\\
F^S_{_{\pm}\xi}(\hat z,\hat n)&=q^{_{\pm}\xi,S} Y^\star_{1 0}(\hat n)\,,\\
F^S_\kappa(\hat z,\hat n)&=q^{\kappa,S} Y^\star_{1 0}(\hat n) \,,\nonumber\\
F^S_{_{\pm}\zeta}(\hat z,\hat n)&=q_0^{_{\pm}\zeta,S} Y^\star_{0 0}(\hat n)+q_1^{_{\pm}\zeta,S} Y^\star_{1 0}(\hat n)\,,\nonumber
\end{align}
with
\begin{align}
q^{z,S}_\ell&=\sqrt{\pi}\left(-\sqrt{3}\right)^{-\ell}\,,\nonumber\\
q^{_{\pm}\xi,S}&=\sqrt{\frac{\pi}{3}}\,,\\
q^{\kappa,S}&=\sqrt{\frac{\pi}{3}}\,,\nonumber\\
q_\ell^{_{\pm}\zeta,S}&=\mp\left(-\sqrt{3}\right)^\ell\sqrt{\pi}\,.\nonumber
\end{align}
For what concerns the ${\,_{\pm}\xi}$ and the ${\,_{\pm}\zeta}$ coefficients, since the inverse operations, necessary to recover the original quantities are not defined for $\ell=0$, we cannot raise the spin and proceed as in \cite{Mentasti:2024fgt}, therefore we can only deal with the observables for $z$ and $\kappa$. This is a more general result that holds when trying to obtain a diagonal expression in the GW ($\hat q$) and object ($\hat n$) coordinates in the case where the spin index in the object's coordinates is greater than the spin of the GW considered. 
This result is not surprising. Symmetry arguments preclude a GW sourcing term generating an observable signature when the GW helicity (or spin-weight) is lower than that of an observable that is an irreducible representation under rotations over the two-sphere. 

In the general reference frame, the response functions for $z$ and $\kappa$ read
\begin{align}
F^S_z(\hat q,\hat n)&=\sum_{\ell=0}^1\sum_{m=-\ell}^\ell A^{z,S}_\ell Y^{\,}_{\ell m}(\hat q)Y^\star_{\ell m}(\hat n)\,,\nonumber\\
F^S_\kappa(\hat q,\hat n)&=\sum_{m=-1}^1A^{\kappa,S}_1 Y^{\,}_{\ell m}(\hat q)Y^\star_{\ell m}(\hat n)\,,\nonumber\\
A^{z,S}_\ell&=(-1)^\ell\frac{2\pi}{2\ell+1}\,,\nonumber\\
A^{\kappa,S}_1&=\frac{2\pi}{3}\,.
\end{align}
Since the observables are now defined as real scalar quantities and, contrary to the Einsteinian case, there is only one polarisation for the scalar and longitudinal GWs, the overlap functions introduced in \eqref{eq:gamma_unpol} take the form
\begin{align}\label{eq:ORF_scalar}
\Gamma_{S}^{A,B}(\beta)&=\sum_{\ell}\frac{2\ell+1}{4\pi}\,C_\ell^{AB}\,d_{s's}^\ell(\beta)\,,
\end{align}
where the coefficients $C_\ell$ are shown in Table~\ref{tab:coefficients}.

\subsubsection{Another astrometric observable}

Although, for the case of a scalar GW source, one cannot decompose the astrometric deflection $\delta\hat n$ into the two $_{\pm}\delta$ modes, we can still consider ``building'' an astrometric observable that is sensitive to scalar polarisation. An example is to consider the norm of the astrometric deflection\footnote{Interestingly, the same idea would not work if one considers Einsteinian modes, as the norm of $\delta\hat n$ vanishes. In the case of a longitudinal and a vectorial mode, the norm would not vanish, but the computation of the response functions would not be possible, as it turns out that they cannot be expanded into spherical harmonics when evaluated in the reference frame with the GW coming from the $\hat z$ axis, indicating the presence of pathological divergences. We explore this below.}
\begin{align}
|\delta\hat n^S(\hat z,\hat n)|=\delta\hat n^S(\hat z,\hat n)\cdot \hat b_\theta(\hat q,\hat n)=\frac{1}{2}\sin\theta_s\,.
\end{align}
In this case, we can expand the observable in salary spherical harmonics and proceed as usual to obtain a diagonal spectral function, obtaining
\begin{align}
|\delta\hat n^S(\hat z,\hat n)|&=\sum_{\ell=0}^\infty q^{|\delta \hat n|,S}_\ell Y_{\ell 0}(\hat n)\,,\nonumber\\
q^{|\delta \hat n|,S}_\ell&=2\pi\int d\theta_s \frac{1}{2}\sin\theta_s\sin\theta_s Y_{\ell 0}(\theta_s)=2\pi\sqrt{\frac{2\ell+1}{4\pi}}\int_{-1}^1 d\mu \frac{1}{2}\sqrt{1-\mu^2} P_\ell(\mu)\,,\nonumber\\
&=-\frac{\pi}{4}\sqrt{\frac{2\ell+1}{4\pi}}\begin{cases}
    \frac{\Gamma(\frac{\ell-1}{2})\Gamma(\frac{\ell+1}{2})}{\Gamma(\frac{\ell+2}{2})\Gamma(\frac{\ell+4}{2})}\\
    0\qquad \ell\;\text{odd}
\end{cases}\,,
\end{align}
where $\Gamma(z)$ is Euler's Gamma function. In the most general reference frame, we find
\begin{align}\label{eq:norm_diagonalized}
|\delta\hat n^S(\hat q,\hat n)|&=\sum_{\ell m}A^{|\delta \hat n|,S}_\ell Y_{\ell m}(\hat q)Y^\star_{\ell m}(\hat n)\,,\nonumber\\
A^{|\delta \hat n|,S}_\ell&=-\frac{\pi}{4}\begin{cases}
    \frac{\Gamma(\frac{\ell-1}{2})\Gamma(\frac{\ell+1}{2})}{\Gamma(\frac{\ell+2}{2})\Gamma(\frac{\ell+4}{2})}\\
    0\qquad \ell\;\text{odd}
\end{cases}\,.
\end{align}
Note that for $\ell\gg 1$ the coefficients scale as $A^{|\delta \hat n|,S}_\ell\simeq-2/\ell^3$, therefore the series that defines the spherical harmonics expansion in Eq.~\eqref{eq:norm_diagonalized} converges to a finite function for every value of $\hat q$ and $\hat n$.
\begin{figure}[t!]
    \centering
    \begin{subfigure}[b]{0.49\textwidth}
        \centering
        \includegraphics[width=\textwidth]{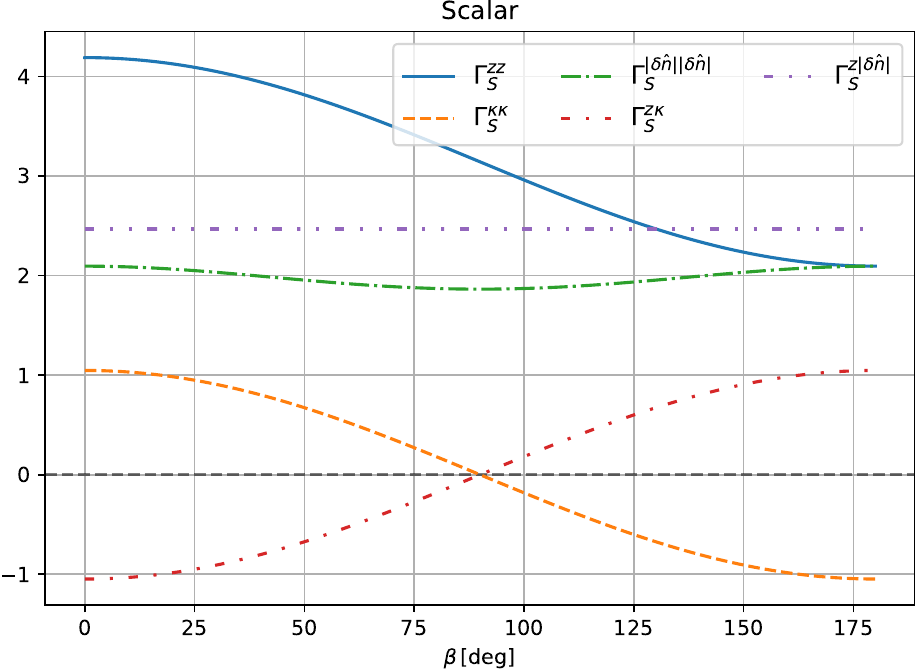}
    \end{subfigure}
    \hfill
    \begin{subfigure}[b]{0.5\textwidth}
        \centering
        \includegraphics[width=\textwidth]{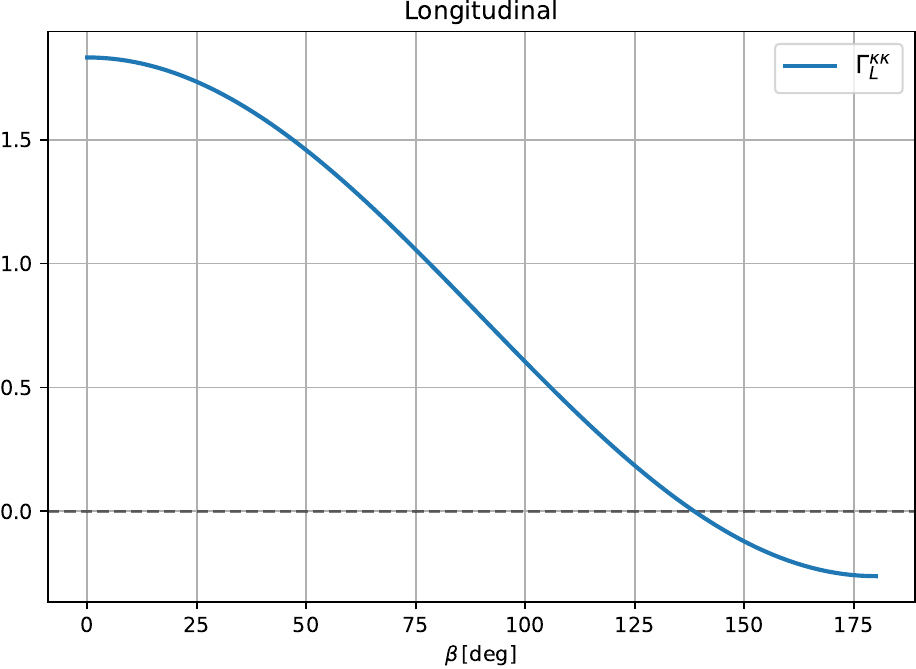}
    \end{subfigure}
    \caption{Angular correlation functions for an Einsteinian-unpolarized GW background analogous to the one defined in \eqref{eq:Gamma_unpol_def} for a scalar/breathing-polarised GW background (left) and a longitudinal-polarised one (right). The correlation functions are evaluated in the reference frame where $\hat n_1=\hat z$ as a function of $\beta=\arccos(\hat n_1\cdot\hat n_2)$. As explained in the text, these are the only observables whose correlators can be computed consistently.}
    \label{fig:Scalar}
\end{figure}
In Fig.~\ref{fig:Scalar} we show the overlap functions of \eqref{eq:ORF_scalar} for the astrometric and redshift observables.

\subsection{Longitudinal polarisation response and spectral functions}
Similarly to what has been done in the previous sections, we evaluate all the observables considered for the case of a purely longitudinal mode (please note that in this case, the longitudinal mode only has one polarisation, unlike the Einsteinian case). A straightforward calculation of the responses in the aligned frame gives \eqref{eq:distortion_E}:
\begin{align}\label{eq:vars_L}
F^L_{z}(\hat z,\hat n)&=\frac{\cos^2\theta_s}{2(1+\cos\theta_s)}\,,\nonumber\\
F^L_{\,_{\pm}\delta}(\hat z,\hat n)&=\frac{\cos\theta_s\sin\theta_s}{2(1+\cos\theta_s)}\,,\nonumber\\
F^L_{\kappa}(\hat z,\hat n)&=\frac{1}{4}(1-2\cos\theta_s)\,,\\
F^L_{\omega}(\hat z,\hat n)&=0 \,,\nonumber\\
F^L_{{}_{\pm}\gamma}(\hat z,\hat n)&=\frac{1}{4}\tan^2\left(\frac{\theta_s}{2}\right)=\frac{1}{4}\frac{1-\cos\theta_s}{1+\cos\theta_s}\,.\nonumber
\end{align}
At this point, one would proceed to expand the redshift response functions in spherical harmonics $Y_{\ell 0}(\hat n)$ as we did in the Einsteinian case. However, this is not possible, as the redshift response function in \eqref{eq:vars_L} is square integrable over the two-sphere due to the divergence. This indicates that the response function cannot be expanded in harmonic space. Indeed, the redshift overlap function cannot be computed even numerically.
For the astrometric and the shimmering observables, we find that, as happened for the scalar/breathing mode in the previous section, the $_{\pm}\delta$ and $_{\pm}\gamma$ coefficients cannot be computed consistently using the spin-raising (-lowering) method because the observables have higher helicity than the GW that sources them.  Therefore, we are left with only the $\kappa$ coefficient, whose response function evaluates to, after a set of steps similar to the ones done in the previous sections,
\begin{align}
F^L_\kappa(\hat q,\hat n)&=\sum_{\ell m}A^{\kappa,L}_\ell Y_{\ell m}(\hat q)Y^\star_{\ell m}(\hat n)\,,\nonumber\\
A^{\kappa,L}_\ell&=\sqrt{\frac{4\pi}{2\ell+1}}q_\ell^{\kappa,L}=\frac{\pi}{\sqrt{2\ell+1}}\left(-\frac{2}{\sqrt{3}}\right)^\ell\,, \qquad \mbox{for}\,\, \ell=\{0,1\}\,.
\end{align}

\subsection{Vectorial polarisation response and spectral functions}
The computation of the response and the overlap functions for the vectorial GW polarisations is very similar to the Einsteinian case, as there are two polarisations present ($x,y$), analogous to the two Einsteinian canonical polarisations ($+,\times$).
We introduce the \textit{vectorially-left} polarised GW 
\begin{equation}
    h_{ij}^{L}(\hat q)=h_{ij}^x(\hat q)+i\;h_{ij}^y(\hat q)\,,
\end{equation}
and the response functions for the redshift to a vectorially left-polarised GWs can be obtained by noticing that 
\begin{equation}
    h^{L,v}_{ij}(\hat q)=\hat v_i(\hat q)\hat q_j+\hat q_i\hat v_j(\hat q)\,,
\end{equation}
with $\hat v(\hat q)=\hat e^\theta(\hat q)+i\hat e^\phi(\hat q)$. Using this, we obtain
\begin{align}\label{eq:distortion_V}
F_z^V (\hat q,\hat n) &= \frac{\left[\hat n\cdot \hat v(\hat q)\right]\left[\hat n\cdot \hat q\right]}{1+\hat q_l\,\hat n^l}\,,\nonumber\\
\delta \hat n^V_i (\hat q,\hat n) &= \frac{1}{2}\left[ \frac{\hat n_i+\hat q_i}{1+\hat q_l\,\hat n^l}2\left[\hat n\cdot \hat v(\hat q)\right]\left[\hat n\cdot \hat q\right] - \left[\hat n\cdot \hat v(\hat q)\right]\hat q_i- \left[\hat n\cdot \hat q\right]\hat v_i(\hat q)\right]\,,\\
\widetilde \psi_{ij}^V (\hat q,\hat n)&= \frac{1}{2}\left[\delta_{ij}- \frac{\hat n_i+\hat q_i}{1+\hat q_l\,n^l}\hat q_j\right] \frac{2\left[\hat n\cdot \hat v(\hat q)\right]\left[\hat n\cdot\hat q\right]}{1+\hat q_l\,\hat n^l}+\frac{\hat n_i+\hat q_i}{1+\hat q_l\,\hat n^l}\left[\hat n\cdot \hat v(\hat q)\right] \hat q_j+\nonumber\\
&\frac{\hat n_i+\hat q_i}{1+\hat q_l\,\hat n^l}\left[\hat n\cdot \hat q\right] \hat v_j(\hat q)-\frac{1}{2} \hat v_i(\hat q)\hat q_j-\frac{1}{2} \hat q_i \hat v_j(\hat q)\,.\nonumber
\end{align}
The redshift, astrometry and \textit{shimmering} response functions in the reference frame with the GW coming from the $z$-axis evaluate to
\begin{align}\label{eq:vars_V}
F^V_{z}(\hat z,\hat n)&=e^{-i(\phi-\phi_s)}\frac{\cos\theta_s\sin\theta_s}{1+\cos\theta_s}\,,\nonumber\\
F^V_{\,_{+}\delta}(\hat z,\hat n)&=\frac{1}{2}e^{-i(\phi-\phi_s)}(1-3\cos\theta_s)\,,\nonumber\\
F^V_{\,_{+}\delta}(\hat z,\hat n)&=\frac{1}{2}e^{-i(\phi-\phi_s)}(1-\cos\theta_s)\,,\nonumber\\
F^V_{\kappa}(\hat z,\hat n)&=\frac{1}{2}e^{-i(\phi-\phi_s)}\left[\tan\left(\frac{\theta_s}{2}\right)-2\sin\theta_s\right]\,,\nonumber\\
F^V_{\omega}(\hat z,\hat n)&=-\frac{i}{2} e^{-i(\phi-\phi_s)}\frac{\cos\theta_s\sin\theta_s}{1+\cos\theta_s}\,,\\
F^V_{{}_{+}\gamma}(\hat z,\hat n)&=0\,,\nonumber\\
F^V_{{}\,_{-}\gamma}(\hat z,\hat n)&=-e^{-i(\phi-\phi_s)}\tan\left(\frac{\theta_s}{2}\right)\,.
\end{align}
Once again, where required, we lower or raise the spin of the responses to obtain scalar quantities
\begin{align}
F^V_{_{+}\xi}(\hat z, \hat n)&\equiv {\slashed \partial}F^V_{{}_{+}\delta}=e^{-i(\phi-\phi_s)}\left(2\tan\frac{\theta_s}{2}-3\sin\theta_s\right)\,,\nonumber\\
F^V_{_{-}\xi}(\hat z, \hat n)&\equiv\bar{\slashed \partial}F^V_{{}_{+}\delta}=-e^{-i(\phi-\phi_s)}\sin\theta_s\,,\\
F^V_{_{-}\zeta}(\hat z, \hat n)&\equiv{\slashed \partial}^2F^V_{{}\,_{-}\gamma}=\left(-\partial_\mu-\frac{1}{1-\mu^2}\right)^2\left[\left(1-\mu^2\right) F^V_{{}\,_{-}\gamma}\right]=e^{-i(\phi-\phi_s)}2\sqrt{\frac{1-\cos\theta_s}{1+\cos\theta_s}} \,.\nonumber
\end{align}
The scalar response functions are expanded as
\begin{align}
F^V_{z}(\hat z,\hat n)&=\sum_{\ell=1}^\infty q_\ell^{z,V} Y^\star_{\ell -1}(\hat n) \,, \nonumber\\
F^V_{_{\pm}\xi}(\hat z,\hat n)&=\sum_{\ell=1}^\infty q_\ell^{_{\pm}\xi,V} Y^\star_{\ell -1}(\hat n)\,, \nonumber\\
F^V_\kappa(\hat z,\hat n)&=\sum_{\ell=1}^\infty q_\ell^{\kappa,V} Y^\star_{\ell -1}(\hat n) \,, \\
F^V_\omega(\hat z,\hat n)&=\sum_{\ell=1}^\infty q_\ell^{\omega,V} Y^\star_{\ell -1}(\hat n) \,,\nonumber\\
F^V_{_{-}\zeta}(\hat z,\hat n)&=\sum_{\ell=1}^\infty q_\ell^{_{-}\zeta,V} Y^\star_{\ell -1}(\hat n)\,, \nonumber
\end{align}
with
\begin{align}
q_\ell^{z,V}&=(-1)^{\ell}\sqrt{\frac{4\pi(2\ell+1)}{(\ell+1)\ell}}\left(1-\frac{2}{3}\delta_{\ell 1}\right)e^{-i\phi}\,,\nonumber\\
q_\ell^{_{+}\xi,V}&=-2(-1)^\ell\sqrt{\frac{4\pi(2\ell+1)}{(\ell+1)\ell}}\left(1\,_{-}\delta_{\ell 1}\right)e^{-i\phi}\,,\nonumber\\
q_\ell^{_{-}\xi,V}&=\frac{2}{3}\sqrt{\frac{4\pi(2\ell+1)}    {(\ell+1)\ell}}\delta_{\ell 1} e^{-i\phi}\,,\\
q_\ell^{\kappa,V}&=(-1)^{\ell+1}\sqrt{\frac{\pi(2\ell+1)}{(\ell+1)\ell}}\left(1-\frac{4}{3}\delta_{\ell 1}\right)e^{-i\phi}\,,\nonumber\\
q_\ell^{\omega,V}&=i(-1)^{\ell+1}\sqrt{\frac{\pi(2\ell+1)}{(\ell+1)\ell}}\left(1-\frac{2}{3}\delta_{\ell 1}\right)e^{-i\phi}\,,\nonumber\\
q_\ell^{_{-}\zeta,V}&=4(-1)^{\ell+1}\sqrt{\frac{\pi(2\ell+1)}{(\ell+1)\ell}}e^{-i\phi}\,.\nonumber
\end{align}

\begin{figure}[t!]
  \centering
    \centering
\includegraphics[width=10cm]{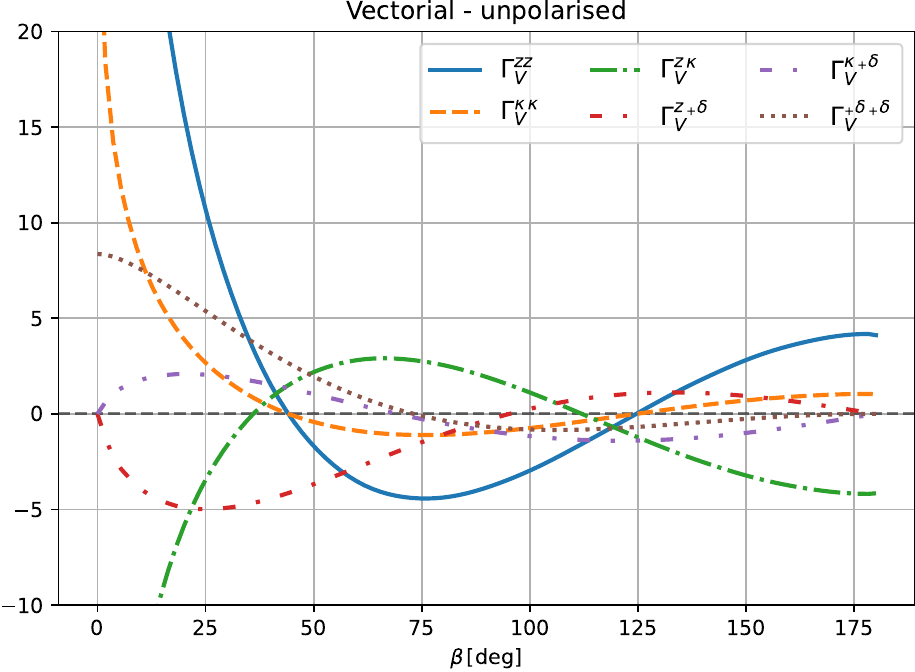}
\caption{%
    Angular correlation functions for a vectorial‑unpolarised GW background, defined in Eq.~\eqref{eq:Gamma_unpol_def}, in the aligned frame. We show only unique functions and list the equivalent ones in Table~\ref{tab:Vectorial-symmetries}}
    \label{fig:Vectorial}
\end{figure}

Similar to what we determined for the scalar GW cases, in the vectorial GW case, the tensorial (shimmering) observables, i.e. the ${\,_{\pm}\gamma}$ coefficients, are not sourced. This is manifest in the fact that since the inverse operations, necessary to recover the original quantities, are not defined for $\ell=1$, we cannot raise the spin and proceed as in \cite{Mentasti:2024fgt}. Once again, this confirms the result that we should expect simply from symmetry arguments. Therefore, in this case, we only need to consider the case of $\kappa$, $\omega$, and ${\,_{\pm}\delta}$.
In the general frame, these read
\begin{align}
F^V_{z}(\hat q,\hat n)&=\sum_{\ell m}A^{z,V}_\ell \,_{1}Y_{\ell m}(\hat q)Y^\star_{\ell m}(\hat n)\,,\nonumber\\
F^V_{\,_{+}\delta}(\hat q,\hat n)&=\sum_{\ell m}A^{\,_{+}\delta,V}_\ell \,_{1}Y_{\ell m}(\hat q)\,_{1}Y^\star_{\ell m}(\hat n)\,,\nonumber\\
F^V_{\,_{+}\delta}(\hat q,\hat n)&=\sum_{\ell m}A^{\,_{+}\delta,V}_\ell \,_{1}Y_{\ell m}(\hat q)\,_{-1}Y^\star_{\ell m}(\hat n)\,,\\
F^V_\kappa(\hat q,\hat n)&=\sum_{\ell m}A^{\kappa,V}_\ell \,_{1}Y_{\ell m}(\hat q)Y^\star_{\ell m}(\hat n)\,,\nonumber\\
F^V_{\omega}(\hat q,\hat n)&=\sum_{\ell m}A^{\omega,V}_\ell \,_{1}Y^{\,}_{\ell m}(\hat q)Y^\star_{\ell m}(\hat n)\,,\nonumber
\end{align}
with diagonal spectral functions
\begin{align}
A^{z,V}_\ell&=-(-1)^{\ell}\frac{4\pi}{\sqrt{(\ell+1)\ell}}\left(1-\frac{2}{3}\delta_{\ell 1}\right)\,,\nonumber\\
A^{\,_{+}\delta,V}_\ell&=(-1)^\ell\frac{8\pi}{(\ell+1)\ell}\left(1\,_{-}\delta_{\ell 1}\right)\,,\nonumber\\
A^{\,_{-}\delta,V}_\ell&=-\frac{4\pi}{3}\delta_{\ell 1}\,,\\
A^{\kappa,V}_\ell&=(-1)^{\ell}\frac{2\pi}{\sqrt{(\ell+1)\ell}}\left(1-\frac{4}{3}\delta_{\ell 1}\right)\,,\nonumber\\
A^{\omega,V}_\ell&=i(-1)^{\ell}\frac{2\pi}{\sqrt{(\ell+1)\ell}}\left(1-\frac{2}{3}\delta_{\ell 1}\right)\,.\nonumber
\end{align}
\begin{table}
    \centering
    \begin{tabular}{|l | l |} 
     \hline
     Prototype & Equivalent overlap functions \\ [0.5ex] 
     \hline\hline
     $\Gamma^{zz}$ & $4\,\Gamma^{\omega\omega}$ \\ 
     \hline
     $\Gamma^{z\kappa}$ & $\Gamma^{\kappa z}$ \\ 
     \hline
     $\Gamma^{z_{+}\delta}$ & $\Gamma^{_{-}\delta z}$, $\,-\Gamma^{z_{-}\delta}$, $\,-\Gamma^{_{+}\delta z}$, $-2i\,\Gamma^{\omega_{+}\delta}$, $-2i\,\Gamma^{_{-}\delta\omega}$, $-2i\,\Gamma^{_{+}\delta\omega}$, $-2i\,\Gamma^{\omega_{-}\delta}$ \\ 
     \hline
     $\Gamma^{\kappa_{+}\delta}$ & $\Gamma^{_{-}\delta\kappa}$, $\,-\Gamma^{\kappa_{-}\delta}$, $\,-\Gamma^{_{+}\delta\kappa}$ \\ \hline
     $\Gamma^{_{+}\delta_{+}\delta}$ & $\Gamma^{_{-}\delta_{-}\delta}$\\
     \hline
    \end{tabular}
  \caption{
     Lookup table of equivalent angular correlation functions for a vectorial‑unpolarised GW background.\label{tab:Vectorial-symmetries}}
\end{table}
In Fig.~\ref{fig:Vectorial} we display the overlap functions between astrometric and redshift observables for a vectorially-unpolarised background obtained from \eqref{eq:gamma_unpol} in a completely analogous way to what has been done in the Einsteinian case
\begin{align}\label{eq:Vcombine}
\Gamma_{V}^{A,B}(\beta)&=\frac{1}{2}\left[\Gamma_{v,V}^{A,B}(\beta)+\Gamma_{v,V}^{\bar B,\bar A}(\beta)\right]\,,
\end{align}
where
\begin{align}\label{eq:CellFinal}
\Gamma_{v,V}^{AB}(\beta) &=\frac{1}{4\pi}\sum_{\ell}\,(2\ell+1)\,C_\ell^{AB}\,d_{s's}^\ell(\beta)\,.
\end{align}
with the coefficients $C_\ell^{AB}$ in Table~\ref{tab:coefficients}.

\begin{table}[h!]
\centering
\begin{tabular}{|c | c | c | c | c | c | c |} 
 \hline
 & Einsteinian ($s_{
 \rm GW}=2)$& Vectorial ($s_{
 \rm GW}=1)$& Breathing ($s_{
 \rm GW}=0)$& Longitudinal ($s_{
 \rm GW}=0)$& $s$ & $s'$ \\ [0.5ex] 
 \hline\hline
 $4 \pi^2\,C_\ell^{\kappa\kappa}$ & $4\,_2^{\,}{\cal N}_\ell^2$ & $\, _1^{\,}{\cal N}_\ell^2\left(1-\frac{8}{9}\delta_{\ell1}\right)$ & $\frac{1}{9}\delta_{\ell1}$ & $\frac{1}{4}\left(\frac{4}{3}\right)^\ell/(2\ell+1)$ & 0 & 0 \\ 
 \hline
 $4 \pi^2\,C_\ell^{z\kappa}$ & $4\,_2^{\,}{\cal N}_\ell^2$ & $-2 \, _1^{\,}{\cal N}_\ell^2\left(1-\frac{10}{9}\delta_{\ell1}\right)$ & $-\frac{1}{9}\delta_{\ell 1}$ & - & 0 & 0 \\ 
 \hline
$4 \pi^2\, C_\ell^{zz}$ & $4 \,_2^{\,}{\cal N}_\ell^2$ & $4\, _1^{\,}{\cal N}_\ell^2\left(1-\frac{8}{9}\delta_{\ell1}\right)$ & $(\delta_{\ell 0}+\delta_{\ell 1})/(2\ell+1)^2$ & - & 0 & 0 \\ 
 \hline
 $4 \pi^2\,C_\ell^{\kappa\delta}$ & $-16\,_2^{\,}{\cal N}^2_\ell/\sqrt{\ell(\ell+1)}$ & $4\, _1^{\,}{\cal N}_\ell^3\left(1-\delta_{\ell1}\right)$ & - & - & 0 & -1 \\ 
 \hline
$4 \pi^2\,C_\ell^{z\delta}$ & $-16\,_2^{\,}{\cal N}^2_\ell/\sqrt{\ell(\ell+1)}$ & $-8\, _1^{\,}{\cal N}_\ell^3\left(1-\delta_{\ell1}\right)$ & - & - & 0 & -1 \\ 
 \hline
 $4 \pi^2\,C_\ell^{\delta\delta}$ & $64\,_2^{\,}{\cal N}^2_\ell/\ell(\ell+1)$ & $16 _1^{\,}{\cal N}_\ell^4 \left(1-\delta_{\ell1}\right)$ & - & - & -1 & -1 \\
 \hline
$4 \pi^2\,C_\ell^{z\gamma}$ & $8\,_1^{\,}{\cal N}_\ell^2\,_2^{\,}{\cal N}_\ell$ & - & - & - & 0 & 2 \\
 \hline
$4 \pi^2\,C_\ell^{\kappa\gamma}$ & $8\,_1^{\,}{\cal N}_\ell^2\,_2^{\,}{\cal N}_\ell$ & - & - & - & 0 & 2 \\ 
 \hline
$4 \pi^2\,C_\ell^{\delta\gamma}$ & $-32\,  _1^{\,}{\cal N}^2_\ell \,_2^{\,}{\cal N}_\ell/\sqrt{\ell(\ell+1)}$ & - & - & - & -1 & 2 \\
 \hline
$4 \pi^2\,C_\ell^{\gamma\gamma}$ & $16\,_1^{\,}{\cal N}_\ell^4$ & - & - & - & 2 & 2\\ [1ex] 
 \hline
\end{tabular}
\caption{Summary table showing all unique diagonal spectral functions that can be used to obtain any overlap (angular correlation) function for combination of spin-$s$ and spin-$s'$ observables using Eq.~\eqref{eq:CellFinal} and Tables~\ref{tab:Einsteinian-symmetries} and~\ref{tab:Vectorial-symmetries}. The spectral functions $C_\ell^{AB}$ are the most general, and frame-independent, description of the correlation between observables. The angular correlation functions obtained from them are frame-dependent but the angular phases associated with this dependence can be ignored for practical purposes.\label{tab:coefficients}}
\end{table}

\section{Discussion}\label{sec:discussion}

In this work, we have extended and applied a unified formalism to compute the response and overlap functions associated with three distinct gravitational wave observables: pulsar timing, astrometric deflections, and the recently proposed \textit{shimmering} effect. By expressing all three observables within a common mathematical framework that makes use of the spin-weighted formalism, we demonstrated that the corresponding overlap functions—whether between identical observables or mixed combinations—can be derived in a structurally identical and straightforward manner using a diagonal basis. This unified treatment significantly simplifies the analysis of cross-correlations among these low-frequency gravitational wave detection methods and highlights the deep geometric similarities underlying their respective responses.

The formalism employed here is naturally suited to the use of spin-weighted spherical harmonics, which allow for a compact and transparent decomposition of the observables' angular dependence. Furthermore, we have checked numerically that this approach yields consistent and correct results whenever the spin of the gravitational wave polarisation is greater than or equal to that of the observable. In these cases—such as the standard astrometric deflection vector or the shimmering quadrupole moment—expansion in the appropriate spin-weighted basis proceeds without ambiguity and captures the relevant physical information.

We summarise our results in Table~\ref{tab:coefficients} in terms of angular power spectral functions $C_\ell$ between different combinations of observables. Due to symmetries, there is a large set of equivalences between all possible combinations of observables, and several combinations also vanish due to symmetries, depending on the helicity of the sourcing GWs and/or of the observables being correlated. Our formalism makes these symmetries manifest and easy to track. To obtain any combinations, the unique prototype functions listed in Table~\ref{tab:coefficients} can be combined as in Eq.~\eqref{eq:Vcombine} to obtain general unpolarised correlation functions. Our $C_\ell$ spectra are the most compact and general form for describing the correlations in observables. The $C_\ell$ are frame-independent- all observers would agree on their reconstruction independent of the alignment of their local basis functions used to define the astrometric and shimmering components. This is because the expansion onto spin-weighted spherical harmonics removes the artificial frame-dependence that is inherent when calculating angular correlation functions.~\footnote{Present even in cases where the signal is statistically isotropic.} The choice of working in harmonic or angular space is, however, a practical one. For PTA observations, the angular correlation function, the Hellings-Downs curve in that case, is always shown because of the sparse and uneven sky coverage of the observations, which would induce severe $\ell-\ell'$ correlations in the harmonic domain. This will not be the case for astrometric and shimmering observations, which rely on surveys with large numbers of sources more evenly spaced on the sky. In those cases, presenting data in the harmonic domain will be more advantageous, as is done in CMB and weak lensing cases.
 
However, some subtleties arise when attempting to define observables with spin higher than that of the gravitational wave perturbation. In such cases, the decomposition may be ill-defined or fail to yield the correct physical interpretation. In certain situations, alternative observables can still be constructed—for instance, by forming scalar combinations like the norm of a vectorial deflection—but caution is required. Not every such scalar function will be a suitable candidate for harmonic expansion, and only those that admit a well-defined expansion in terms of spin-weighted spherical harmonics of the appropriate spin index can be meaningfully employed within this framework.

The results presented here underscore the versatility and generality of the formalism, as well as its utility in systematically comparing and combining the responses of different gravitational wave observables. This is particularly relevant in the context of the growing capabilities of both PTA and astrometric experiments. Future PTA data releases are expected to offer improved sensitivity and timing precision, while next-generation astrometric surveys—extending the legacy of Gaia—will push to new levels of angular resolution.

Astrometry, in particular, represents a promising and relatively unexplored observational window in gravitational wave astronomy, with the potential to bridge the gap between PTA and higher-frequency surveys. The shimmering effect introduced in our previous work adds a complementary tool to this expanding arsenal, offering a new class of tensorial observables that can probe time-dependent spacetime distortions. The formalism developed here provides a consistent and powerful approach to studying all these effects in a unified language and paves the way for future investigations into both standard and non-Einsteinian gravitational wave signatures using multi-modal datasets.



\section*{Acknowledgments}

We thank Marc Kamionkowski for insightful discussions at the beginning of the present work. G.~M. acknowledges support from the Imperial College London Schr\"odinger Scholarship scheme and from the Balzan Award offered by New College, Oxford, which supported the visit to Johns Hopkins University, Baltimore.

\bibliography{refs}

\end{document}